# EFFECT OF LOW-FREQUENCY SIGNAL ON NANOSCALE MEMRISTOR DEVICE

**T. D. Dongale [a, *] and S. R. Ghatage [b]**

[a] Computational Electronics & Nanoscience Research Laboratory,
School of Nanoscience and Biotechnology, Shivaji University, Kolhapur 416004, India
[b] Department of Electronics, Gopal Krishna Gokhale College, Kolhapur, 416012, India

* Corresponding Author: tdd.snst@unishivaji.ac.in

**Abstract:** In the present report, we have investigated the effect of the low-frequency signal on nanoscale memristor device. The frequency is varied from 2 Hz to 10 Hz and the corresponding effect on the current-voltage characteristics, time domain state variable, charge-magnetic flux relation, memristance-charge relation, memristance-voltage characteristics and memristance-magnetic flux relation are studied. The results clearly suggested that the frequency of the input stimulus plays an important role in the device dynamics.

## Introduction

The memristor device is considered as a new circuit element along with the resistor, capacitor, and inductor [1]. It is a two terminal electronics device and possesses many interesting properties such as nonlinearity, passivity, and memory property [2-4]. The first mathematical theory of memristor was formulated in the year 1971 by L. Chua [1]. The experimental realization of memristor was appeared in the year 2008 by HP research group [5]. Considering the many interesting properties of memristor or memristive device, it is considered as a potential candidate for the future resistive memory devices [6], neuromorphic computing application [7-9], nonlinear circuits [10-12], and hardware security applications [13].

The modeling and simulation is an important step towards the realization of functional memristor devices for different applications. The simplest model of memristor was developed HP research group by considering the state variable of the device proportional to the applied voltage. This model generally is known as linear drift model of the memristor device [5]. In the present report, we have investigated the effect of the low-frequency signal on nanoscale memristor device. In order to simulate the nanoscale memristor device, we have used linear drift model of the memristor device [5]. The frequency is varied from 2 Hz to 10 Hz and corresponding effect on the current (I)-voltage (V) characteristics, time domain state variable (w), charge (q)-magnetic flux (Φ) relation, memristance (M)-charge (q) relation, memristance (M)-voltage (V) characteristics and memristance (M)-magnetic flux (Φ) relation are studied. The results clearly suggested that the frequency of the input stimulus plays an important role in the device dynamics.

## Computational Details

In the present simulation, we have considered the Pt/$TiO_2$/Pt device structure. The size of the active $TiO_2$ region was 10 nm, furthermore, the size of the doped region was 2 nm. The amplitude of the input sinusoidal signal was 2V and frequency of the signal varied from 2 Hz to 10 Hz. The drift velocity of oxygen vacancies were considered as $\mu_V=10^{-14}$ $m^2V^{-1}s^{-1}$. The value of low resistance state (LRS) or $R_{ON}$ and high resistance state (HRS) or $R_{OFF}$ are kept at 120 Ω and 20 KΩ respectively. The simulation code is scripted in the MATLAB software. The details mathematical formulation can be found in the ref. [14-18].

## Results and Discussion

Fig. 1 represents the Current-Voltage (I-V) characteristics of the Pt/$TiO_2$/Pt memristor device. The frequency is varied from 2 Hz to 10 Hz with step size 2 Hz. It is observed that the area under the curve for low-frequency signal is maximum (up to 4 Hz) and the pinched hysteresis loop tends to decrease as the frequency increases. The obtained results are well matched with the theoretical proposition of memristor device [19]. In order to investigate the frequency dependent I-V relation, we have simulated the state variable dynamics with respect to time for different frequencies, as shown in Fig. 2.

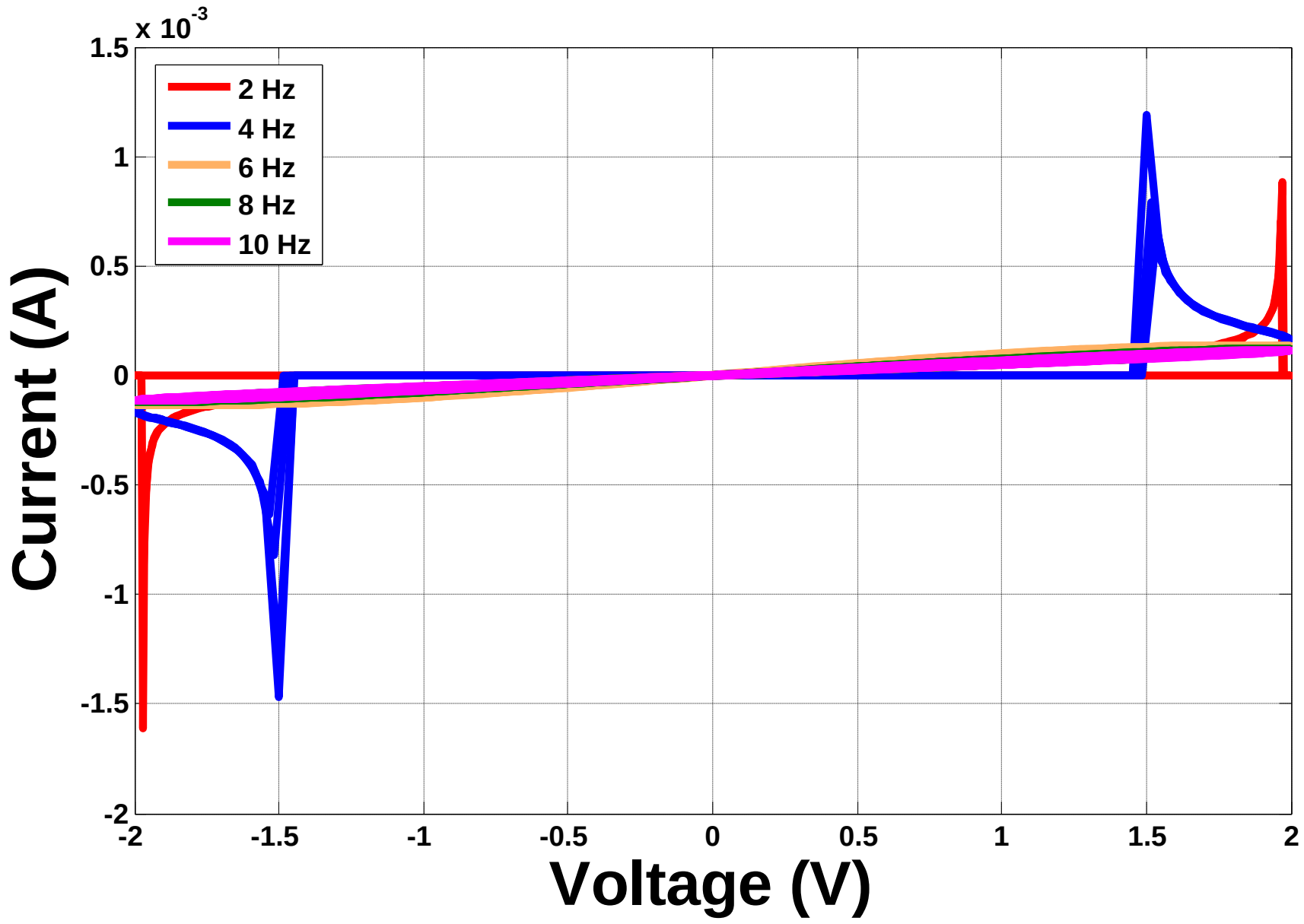

**Fig. 1:** Current-Voltage characteristics of the Pt/$TiO_2$/Pt memristor device. The frequency is varied from 2 Hz to 10 Hz with step size 2 Hz.

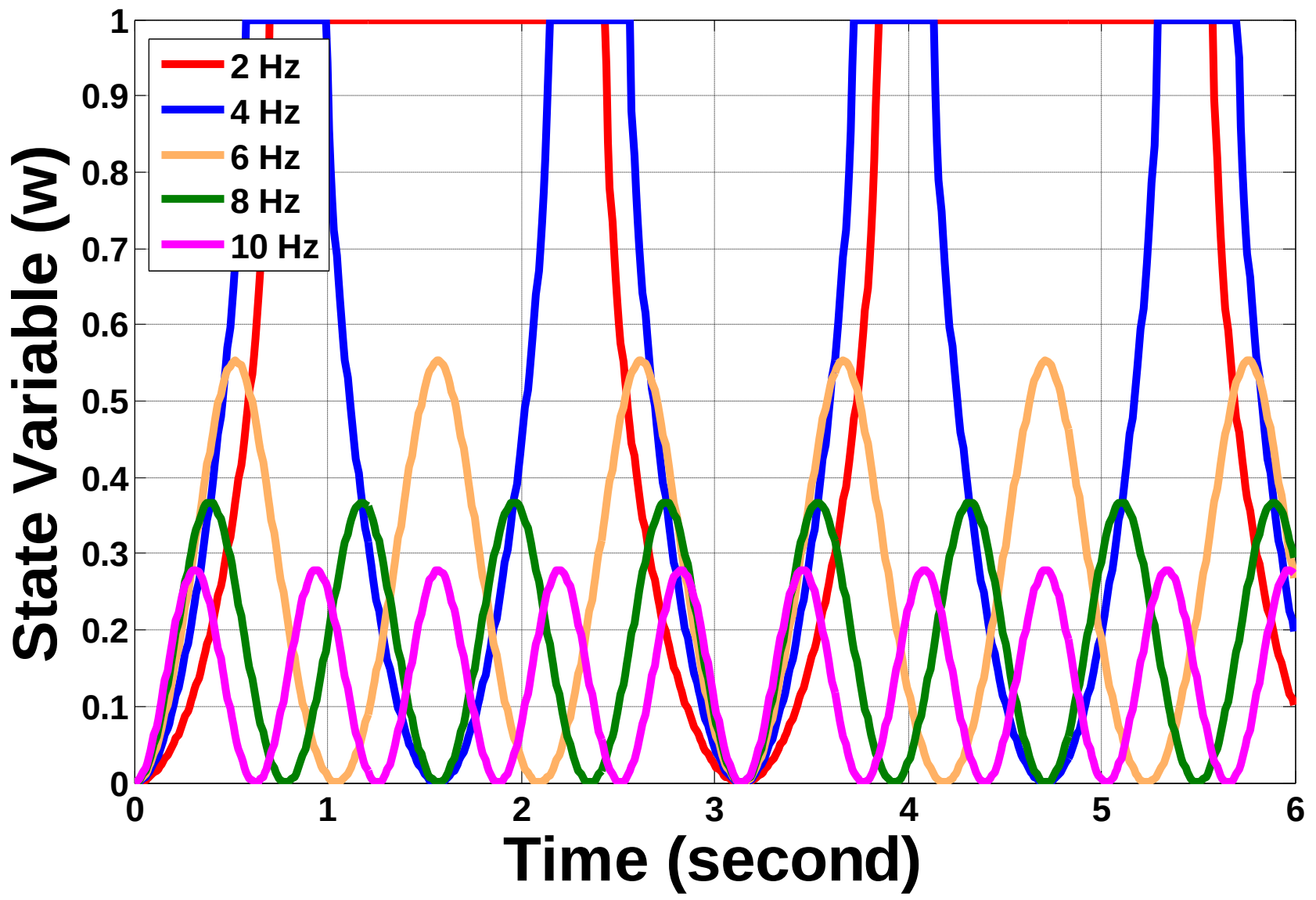

**Fig. 2:** State variable Vs. Time characteristics of the Pt/$TiO_2$/Pt memristor device for different frequencies.

The frequency dependency is originated due to fact that the state variable has enough time to settles its nearest voltage signal for low-frequency signal providing the high area pinched hysteresis loop. In the high-frequency operation, the state variable doesn't have the enough time to settles its nearest voltage signal and this will result in the low area pinched hysteresis loop or linear resistor like characteristics [19-20]. Similarly, the effect of the different frequencies on Charge-Flux (q-Φ) characteristics of the Pt/$TiO_2$/Pt memristor device is shown in Fig. 3 (a-e). It is observed that the charge-flux relation is highly nonlinear at the low frequency (< 4 Hz) and become quasi-linear at high frequency (> 6 Hz).

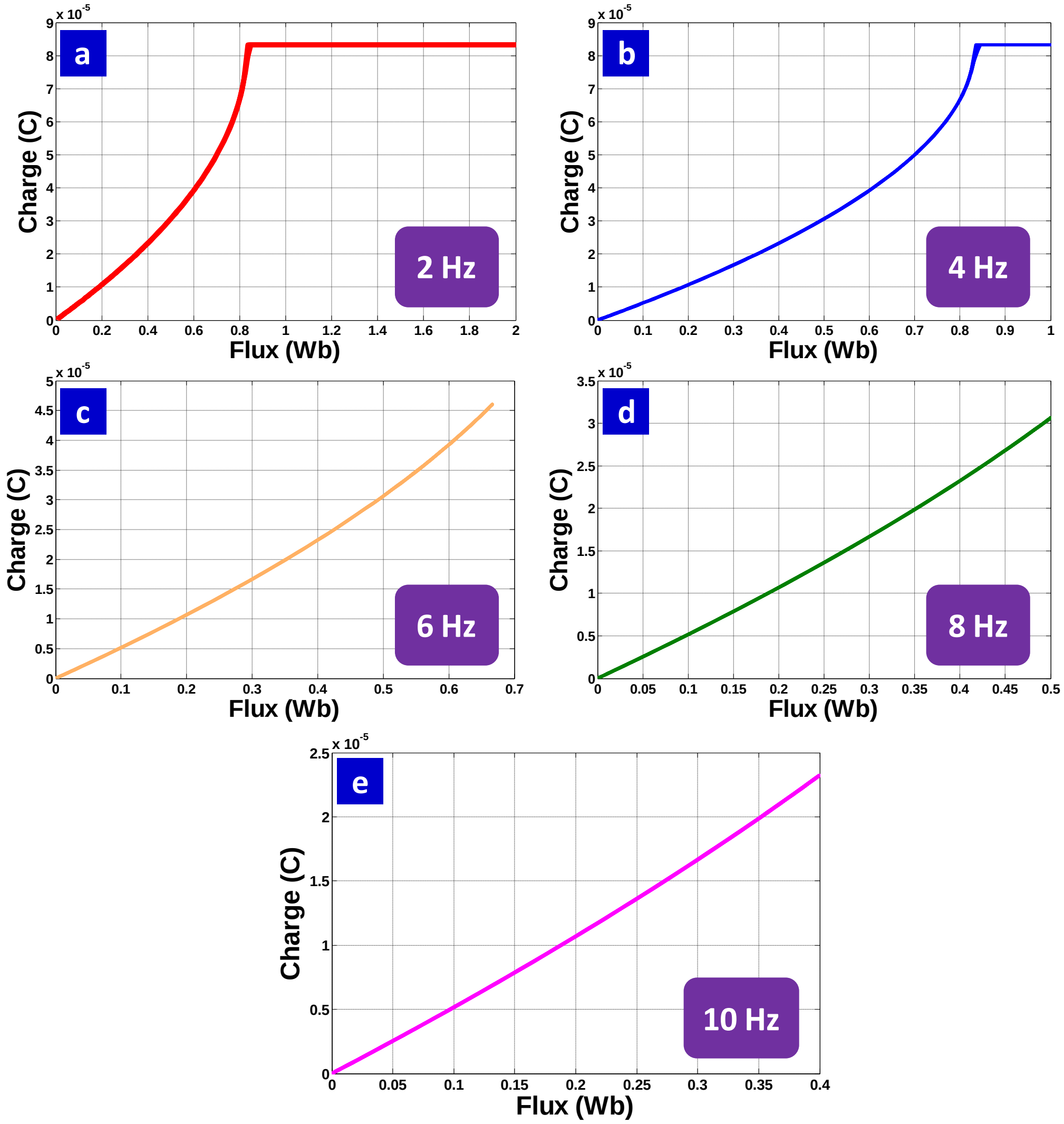

**Fig. 3: (a-e)** Charge-Flux (q-Φ) characteristics of the Pt/$TiO_2$/Pt memristor device for different frequencies.

Furthermore, the Memristance-Charge (M-q) characteristics of the Pt/$TiO_2$/Pt memristor device for different frequencies are shown in Fig. 4. It is observed that the magnitude of the memristance become very high at a lower magnitude of the charge and become low at a higher magnitude of the charge.

This is due to fact that the charge transportation is maximum during the formation of the conductive filament which results in the LRS and becomes minimum during the breaking of the conductive filament which results in the HRS. Moreover, the difference between HRS and LRS is also decreased as the frequency increases. The ratio of HRS to LRS is called as M-factor or memory window [21]. For the practical application, the memory window should be higher to distinguishes the data from the noise. From the results, it is clearly observed that the memory window becomes lower for higher frequency (> 6 Hz). In view of this memristor should be operated at lower frequency input stimulus for the resistive memory application.

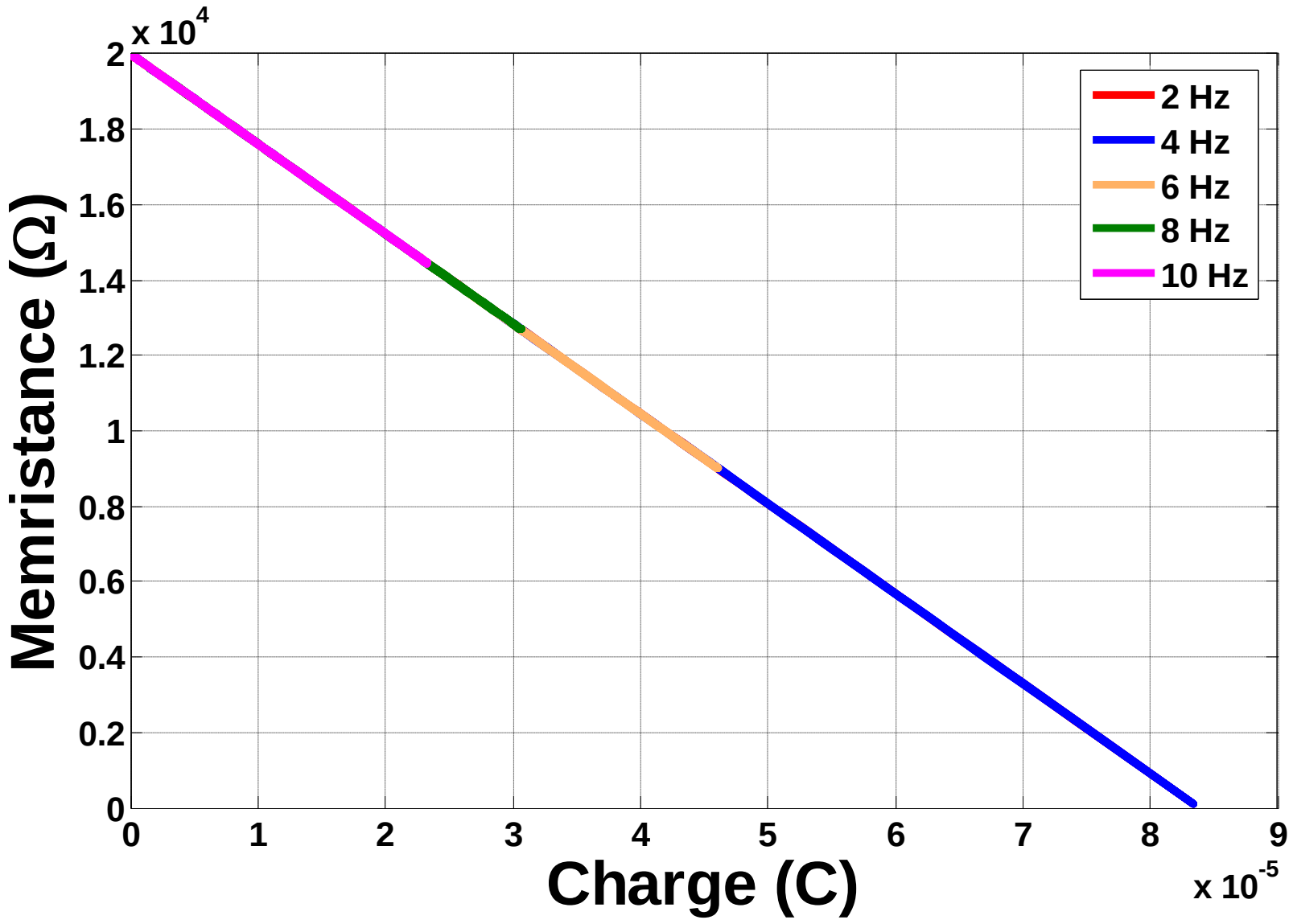

**Fig. 4:** Memristance-Charge (M-q) characteristics of the Pt/$TiO_2$/Pt memristor device for different frequencies.

Two distinct resistance states i.e. LRS and HRS of memristor device can be useful for the development of the high-performance memory devices. One such simulation is depicted in the Fig. 5 (a-e). The results clearly indicate that the two distinct resistance states (LRS and HRS) or abrupt transition from one state to another state are observed only for the low-frequency signals (< 4 Hz) and the abrupt transition vanishes for the high-frequency signals (> 6 Hz). The overall results clearly indicate that the memristor should be operated at the low-frequency region for the better performance.

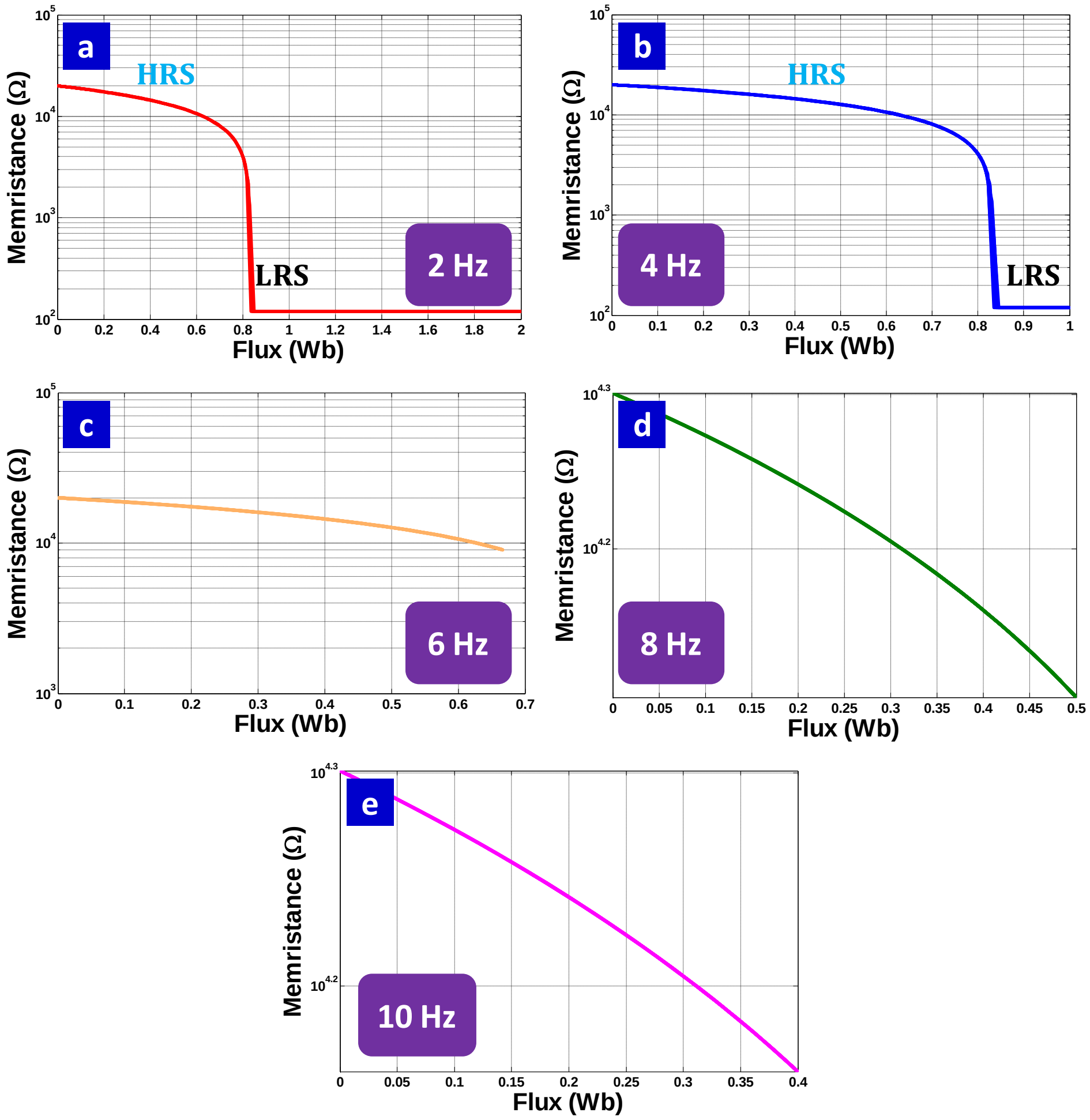

**Fig. 5: (a-e)** Memristance-Flux (M-Φ) characteristics of the Pt/$TiO_2$/Pt memristor device for different frequencies.

## Conclusion

The present simulation investigations reveal some of the important performance parameters of the memristor device. It is found that the pinched hysteresis loop is frequency dependent property and it is dependent on the state variable dynamics. The charge-flux relation of memristor device is highly nonlinear at the low frequency (< 4 Hz) and become quasi-linear at high frequency (> 6 Hz). Furthermore, higher memory window is observed for the lower frequency region and becomes lower at higher frequency. In view of this memristor should be operated at lower frequency input stimulus for the resistive memory application. In addition to this, the abrupt transition from one state to another state is observed only for the low-frequency signals (< 4 Hz) and it vanishes for the high-frequency signals (> 6 Hz). The overall results clearly indicate that the memristor should be operated at the low-frequency region for the better performance.